# Interdependency of Transmission and Distribution Pricing

Sina Parhizi and Amin Khodaei
Department of Electrical and Computer Engineering
University of Denver
Denver, Colorado
sina.parhizi@du.edu, amin.khodaei@du.edu

*Abstract*— **Distribution markets are among the prospect being considered for the future of power systems. They would facilitate integration of distributed energy resources (DERs) and microgrids via a market mechanism and enable them to monetize services they can provide. This paper follows the ongoing work in implementing the distribution market operator (DMO) concept, and its clearing and settlement procedures, and focuses on investigating the pricing conducted by the DMO. The distribution locational marginal prices (D-LMPs) and their relationship with the transmission system locational marginal prices (T-LMPs) are subject of this paper. Numerical simulations on a test distribution system exhibit the benefits and drawbacks of the proposed DMO pricing processes.**

*Index Terms*—**Power distribution, microgrids, power system economics, electricity markets.**

## Nomenclature

| | |
|---|---|
| $a$ | Elements of the bus-line incidence matrix. |
| $b$ | Load benefit. |
| $C_c$ | Customer payments to the DMO. |
| $C_u$ | DMO payment to the ISO. |
| $C_\Delta$ | DMO cost surplus. |
| $f$ | Superscript for fixed loads. |
| $g$ | Index for bid segments. |
| $\mu$ | Penalty factor for the power deviation. |
| $m$ | Index for distribution system buses. |
| $D$ | Load demand. |
| $P^M$ | Power transfer from the main grid. |
| $PD^M$ | Total assigned power from the main grid. |
| $PL$ | Line power flow. |
| $PL^{max}$ | Line power flow limit. |
| $DX$ | The amount of load awarded to each segment of the bid. |
| $PX^{max}$ | Maximum capacity of each bid segment. |
| $r$ | Superscript for responsive loads (proactive customers). |
| $t$ | Index for hours. |
| $\Delta P$ | Power transfer deviation. |
| $\lambda$ | Transmission locational marginal price (T-LMP). |
| $\lambda^D$ | Distribution locational marginal price (D-LMP). |
| $P_t^{pos}$ | Variables used to linearize absolute value. |
| $P_t^{neg}$ | Variables used to linearize absolute value. |

## I. Introduction

HIGH penetration of proactive customers that utilize variety of distributed energy resources (DER), electric vehicles, and flexible loads are frequently mentioned as the characteristics of future power grids [1]–[7]. This growing trend is advocated in order to handle challenges such as growing demand and boost efficiency in the grid operation while meeting the standards put in place by various environmental regulations [8], [9]. Proactive customers in the distribution system are not completely relying on the utility grid to provide their demand; they can self-sustain, in many cases of DER deployment, and even actively participate in the electricity markets and provide several valuable services. In order for proactive customers to be able to realize their full capabilities, to monetize services they provide and to play a role in system pricing and clearing decision-making processes, it is necessary to modernize the existing distribution system operation and enable these entities to participate in the electricity market [10].

The direct control and dispatch of the proactive customers by the independent system operator (ISO) would create several problems such as potential violation of distribution companies' responsibilities and also creating complexities in the market optimization as the participation of proactive customers grows [11], [12]; hence, several proposals have been offered trying to define an entity that establishes an electricity market and facilitates market participation at the distribution level. Distributed System Platform Provider (DSPP) is introduced in New York via the Reforming the Energy Vision program as one of the early efforts in this direction [13]. DSPP is proposed to operate in accord with the electric utility and be in charge of the market operations. It would coordinate with the ISO, customers, and market participants. According to this proposal, distribution system platform (DSP) is a set of functions provided by the utilities to enable broad market participation. Similar to this effort, the Distribution System Operator (DSO) is introduced in California as an entity to be in charge of providing reliable distribution services to the customers and further ensuring predictability for the ISO. It wouldbe coordinating the physical transactions in the transmission-distribution interface but not necessarily the financial aspects of those transactions that [12], [14]–[16] propose can still be among the responsibilities of ISO. The role of the DSO introduced in [17] is similar to that of an ISO but for the distribution system. The DSO would not only be responsible for the reliable operation of the distribution system but also for providing demand response. The DSO would have transactions with the wholesale market at the substation level on one side and proactive customers on the other side. Depending on the extent of the ISO's responsibility in dispatching resources in the distribution system, there would be different levels of the DSO autonomy in operating the distribution system and the degree of the ISO's control over it. The DSO can only act as an aggregator and deliver ISO's dispatch commands or operate a retail market at the distribution level, or something between these two extremes. In [18] the prospective roles for the DSOs are listed including "managing multidirectional flows, organizing auctions, and offering other incentives" in order to minimize the operation costs of the DSO. In [19] control strategies to coordinate several microgrids within a distribution system via a distribution network operator (DNO) is proposed. The DNO can trade energy with the microgrids and the high-voltage transmission system. The problem is formulated as a two-level optimization where microgrids optimize their energy cost at the lower level and the DNO ensures operational constraints at the higher level of optimization. In [20] it is asserted that the utilities should transition their distribution system operation responsibilities to an independent distribution system operator (IDSO) while owning distribution assets. It is argued that operation of the distribution system by the utilities poses a conflict of interest, and as utilities tend to expand their assets it would reduce greater market participation of the proactive customers. The proposed IDSO would be in charge of the system reliability, provide market mechanisms, and optimally schedule the distribution level resources. It would make system management easier for the utilities and determine the value of services provided by proactive customers more objectively.

Most of these proposals, however, are conceptual and lack a detailed rigorous analysis of distribution market operations. It is necessary to clearly define the roles of different parties involved in the distribution system undergoing market-based evolution and further investigate the detailed market processes. Among the concerns that needs to be addressed is the structure of the distribution market operation, the way it is settled [21], and how the procedures for distribution market clearing and settlement are established. The study in [22] points out the necessity of the DSOs to be capable of providing settlements with the ISO for any resource in the distribution system. Studies in [23], [24], furthermore present a marginal pricing method for the DSO considering the congestion problem in the distribution system, and optimizing the social welfare in a system with high penetration of electric vehicles.

The distribution market operator (DMO) is a rather similar entity discussed by authors in [25][26]. The DMO function focused in this work is to facilitate the establishment of market mechanisms in distribution systems and it will be an interface interacting with both the ISO and proactive customers to enable participation of customers in the wholesale market. The DMO receives the demand bids from customers in the distribution system, aggregates them and submits a single aggregated bid to the ISO. After market clearing by the ISO, the DMO divides the assigned power awarded to it between the participated customers. The DMO can be part of the electric utility company or be formed as a separate entity. In either case, it should be an independent operator so as to guarantee the fairness in the market operation. The DMO manages the financial transactions and electric distribution company (EDC) carries out the physical transactions.

The implementation of a distribution market and establishment of the DMO offers several advantages for customers and the system as a whole: (i) Under distribution markets, the proactive customers' demand is set by the DMO and known with certainty on a day-ahead basis, which would enable an efficient control of the peak demand, increase operational reliability, and improve efficiency; (ii) The proactive customers can participate in the electricity market as a player and exchange power with the utility grid and other customers. The DMO would facilitate market participation and coordinate the proactive customers' interactions with the utility grid to minimize the associated operational risks and uncertainties; (iii) There will be a considerable reduction in the required communication in the system as the proactive customers and the ISO only need to communicate with the DMOs. Considering the listed advantages, and many more that will be obtained once these deployments are more widespread, distribution markets can be considered as both beneficial and necessary components in modern power grids which will help accommodate a large penetration of proactive customers. In terms of the distribution market clearing and settlement, the models proposed in [25], [26] were investigated in detail in [27], in which constant and variable power clearing schemes were studied. This paper focuses on how the locational marginal prices in the transmission side of the DMO (i.e., T-LMP) are reflected in the distribution system locational marginal prices (i.e., D-LMP). Furthermore, microgrids will be considered for studies as a representative of proactive customers. Microgrids models, however, can be simplified to model any other type of the proactive customer, such as prosumers or responsive consumers.

The rest of the paper is organized as follows. The formulation for the proposed DMO market clearing and settlement is demonstrated in section III, numerical simulations are presented in section IV, and the paper is concluded in section V.

II. DISTRIBUTION MARKET MODEL

The DMO collects demand bids from the microgrids in the distribution system, creates an aggregated bid, and submits this bid to the ISO. A typical microgrid bid is shown in Fig. 1.

The ISO collects the demand bids (from the DMOs as well as curtailment service provides) and generation bids (from GENCOs), and determines the generation and load schedule. Once the DMO is notified of the ISO's clearing process decisions, it determines the amount of power generation, transfer and demands within the distribution system, and settles the prices and costs among various participants in the market, with the objective of ensuring an optimal operation and a fair settlement.

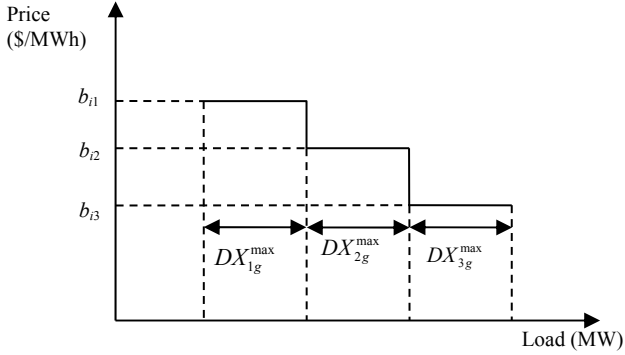

Fig. 1 Demand bid curve for customer at bus *i* with a three-segment bid

### III. DISTRIBUTION MARKET CLEARING AND SETTLEMENT

The DMO's objective is to maximize the distribution system social welfare (1), i.e., load benefits minus generation cost (which is the cost of assigned power from the main grid). This paper proposes to add the last term in the objective to penalize violations in the assigned power.

$$\max \sum_t \sum_m \sum_g b_{mg} DX_{mgt} - \sum_t \lambda_t P_t^M - \sum_t \mu \mid \Delta P_t^M \mid \quad (1)$$

The power assigned to the DMO by the ISO is determined via the wholesale market clearing, hence it is constant. The T-LMP is also determined by the ISO. The penalty coefficient $\mu$ is multiplied with the deviation to ensure that the assigned power will be followed in the distribution network. The objective is subject to distribution network and microgrids prevailing constraints (2)-(6):

$$\sum_l a_{lm} PL_{lt} = D_{mt} \qquad \forall t, \forall m \quad (2)$$

$$\sum_l a_{l0} PL_{lt} - P_t^M = 0 \qquad \forall t \quad (3)$$

$$D_{mt} = \sum_g DX_{mgt} + D_{mt}^f \qquad \forall t, \forall m \quad (4)$$

$$0 \leq DX_{mgt} \leq DX_{mg}^{\max} \qquad \forall t, \forall m, \forall g \quad (5)$$

$$-PL_l^{\max} \leq PL_{lt} \leq PL_l^{\max} \qquad \forall t, \forall l \quad (6)$$

$$\Delta P_t^M = P_t^M - PD_t^M \qquad \forall t \quad (7)$$

$$\Delta P_t^M = P_t^{pos} - P_t^{neg} \qquad \forall t \quad (8)$$

$$\mid \Delta P_t^M \mid = P_t^{pos} + P_t^{neg} \qquad \forall t \quad (9)$$

$$P_t^{neg} \geq 0 \qquad \forall t \quad (10)$$

$$P_t^{pos} \geq 0 \qquad \forall t \quad (11)$$

The nodal power balance is ensured in (2) where the power injected to each bus from connected lines is equal to total bus load. The power balance at transmission-distribution interface is ensured by (3) in which the power transferred by the main grid is distributed among lines connected to this bus (here the bus number 0). The load of passive customers will be constant, while that of proactive customers is variable and defined by the associated segments (4). The scheduled load of microgrids is determined based on the scheduled power consumption in each bid segment, in which each segment is limited by its associated maximum capacity (5). Line power flows are limited by the line capacity limits (6). The added penalty in the objective function, which is represented as the absolute value of the deviation makes the problem nonlinear. In order to linearize this term and to ensure a linear programming problem, (8)-(11) are used. $P_t^{neg}$ and $P_t^{pos}$ are two non-negative variables used to model the absolute value. If the variable inside the absolute value is positive, $P_t^{neg}$ would be equal to zero and when the value is negative $P_t^{pos}$ would be equal to zero. This is guaranteed to happen since the problem is formulated as a linear programming minimization solved by the Simplex method.

The D-LMP in each bus is calculated as the dual variable of the power balance equation in that bus (2), i.e., as a byproduct of the proposed clearing problem. The relationship between the D-LMP and the T-LMP depends on the value of the penalty factor $\mu$. In this paper, the distribution market clearing is conducted in two ways based on the penalty factor: grid-following clearing and grid-independent clearing, as discussed further in the following:

### A. Grid-following clearing

When $\mu=0$ in the proposed formulation, the D-LMP at bus 0 of the distribution (point of connection to the transmission network) will be equal to the T-LMP. In this case the DMO is permitted to import power from the utility grid more/less than the power assigned to it by the ISO, as there would be no penalty. This results in the T-LMP to be reflected in the D-LMPs within the distribution system. At down-stream buses, however, D-LMPs will be determined based on the T-LMP, marginal cost of dispatchable units, and possible distribution line congestions.

### B. Grid-independent clearing

As the amount of $\mu$ is increased, the DMO seeks to minimize the deviation of the scheduled power with the assigned power transfer set by the ISO. In this case the dependency of D-LMPs to the T-LMP will be lowered. For the very large values of the penalty factor, there would be no violation of the power transfer schedule determined by the ISO, while the D-LMPs will be merely functions of the dispatchable units' marginal price and possible distribution line congestions.

## C. Market Settlement

Using D-LMPs, obtained from either methods, the market can be settled, i.e., the payments from customers and the payments to the utility can be determined. The payment of each customer is calculated as the D-LMP times the associated load. The total customer payments is the summation of all payments (12), in which $D_{mt}$ includes both consumers and microgrids. The payment to the utility is calculated as the T-LMP times the assigned power by the ISO (13). Since losses are ignored in the proposed market clearing model, the sum of the distribution loads will be equal to the total power assigned by the ISO (14).

$$C_c = \sum_t \sum_m \lambda_{mt}^D D_{mt} \qquad (12)$$

$$C_u = \sum_t \lambda_t P_t^M \qquad (13)$$

$$P_t^M = \sum_m D_{mt} \qquad \forall t \qquad (14)$$

Considering the payments, the DMO cost surplus can be calculated as the difference between the two calculated payments as in (15):

$$C_\Delta = C_c - C_u = \sum_t \sum_m (\lambda_{mt}^D - \lambda_t) D_{mt} \qquad (15)$$

The obtained $C_\Delta$ can be negative, positive, or zero. The cost settlement is one of the challenges facing DMOs as they operate radial networks, as opposed to the wholesale power system operated by ISOs, and they should guarantee a fair market participation by customers at different locations across the feeders. This issue will be the topic of a future research by authors.

## IV. NUMERICAL RESULTS

The IEEE 13-bus test system [28] in used to investigate the viability and the merits of the proposed processes. Fig. 2 depicts this system in which microgrids are located at buses 2, 3, 5-7, and 10-13. Each customer submits a four-segment power demand bid of maximum 10 MW. A large capacity for distribution lines is considered, however it is assumed that lines 3-8 and 4-5 have smaller capacities and then subject to potential congestions. Various cases have been studied considering the various values for parameters in (1).

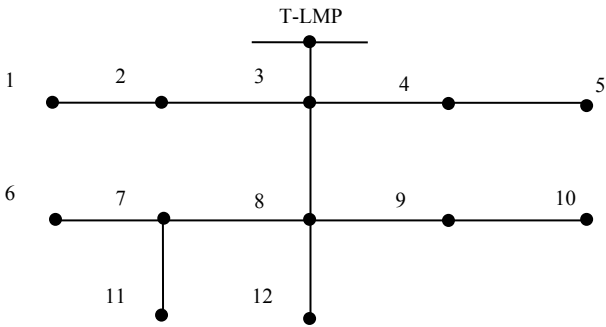

Fig. 2 IEEE 13-bus standard test system.

**Case 1:** In this case $\mu$ is assumed to be 0, and the distribution market clearing results for a variety of T-LMPs is determined. The results of this case are shown in Fig. 3, which illustrates the effect different values of the T-LMP on the marginal price of each bus in the distribution system. A scaling factor is used to change T-LMPs. As the scaling factor increases the power is to be purchased at a higher rate, resulting in a lower power transfer from the ISO and more local generation. This results in lower congestion as the prices of the all busses tend to be equal at higher values of the scaling factor. At lower T-LMP values, D-LMPs follow the T-LMPs and can also impact the grid prices as they respond to the price variations by modifying their power injections. At lower scaling factors, congestion at line 3-8 results in a rise of D-LMPs in buses 6-12. These buses have the same D-LMPs as the lines in the downstream of the feeder do not become constrained.

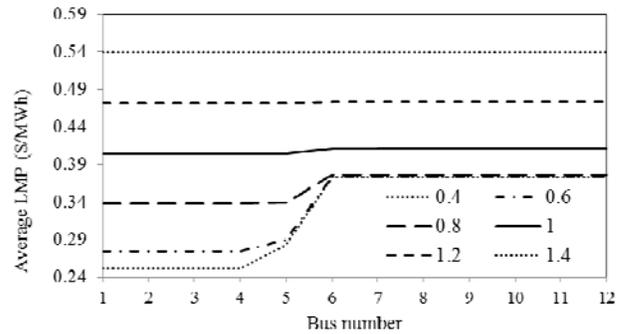

Fig. 3 Daily average LMP at each bus for different T-LMP values (using a scaling factor)

**Case 2:** In this case the second term in (1) is assumed to be 0, i.e., the T-LMP is negligible, and $\mu$ is varied, so the independent operation of the distribution market can be analyzed. The results of this case are shown in Fig. 4, which illustrates the effect of raising $\mu$ on D-LMPs. As $\mu$ increases, the DMO seeks to minimize the deviation from the assigned power transfer. The prices, however, tend to approach the prices when the scheduled power is used without any option to deviate. When $\mu$ approaches infinity, the D-LMPs are functions of marginal prices of the dispatchable units and become independent of the T-LMPs. This would result in the settlement costs of the distribution system be higher, lower, or equal to the payments to the ISO depending on the marginal costs.

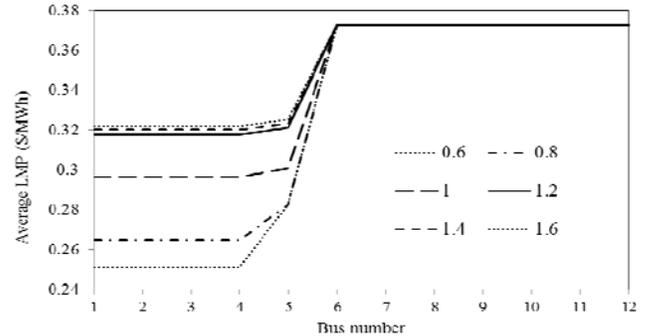

Fig. 4 Daily average D-LMP as a function of $\mu$.

**Case 3:** In this case the T-LMP and $\mu$ are both nonzero. The results of this case are shown in Figs. 5 and 6. In Fig. 5, T-LMP scaling factor varies between 0.1 and 0.9 while $\mu$ is kept at 1. The corresponding difference between the customers' payment to the DMO and payment to the ISO is depicted in Fig. 6 (a deficit for the DMO). It means that while the DMO tries to minimize the deviation of the power transfer with respect to the scheduled power, a term exists in the objective that seeks to minimize the power transfer (even at the expense of higher deviation) to reduce the payments to the ISO.

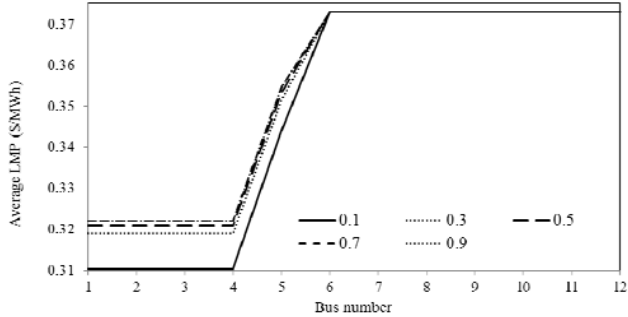

Fig. 5 Daily average D-LMPs as a function of scaling factors while $\mu$ =1.

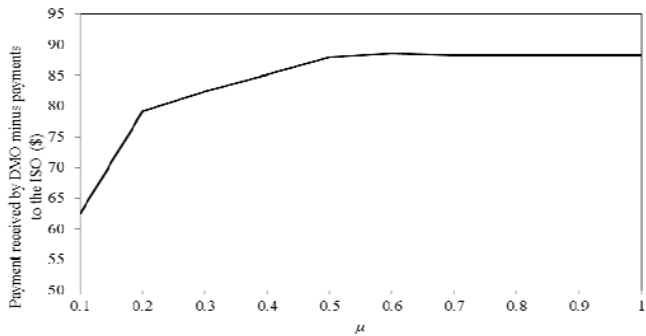

Fig. 6 Payment received by DMO minus payments to the ISO.

## V. CONCLUSION

One of the challenges in the operation of distribution markets is the clearing and settlement method they should use, and the consequent impacts on the distribution market prices. This paper studied the interdependencies between D-LMPs at a distribution network with the T-LMP at the upstream node. It was shown that by adding a penalty factor the D-LMPs could be calculated in a way that the prices follow the associated T-LMP or be determined independent of that. It was further shown that these price would significantly change the market settlement. The decision on the exact value for the penalty coefficient, however, should be made by the DMO.

## VI. REFERENCES

[1] S. Parhizi, H. Lotfi, A. Khodaei, and S. Bahramirad, "State of the Art in Research on Microgrids: A Review," *IEEE Access*, vol. 3, 2015.
[2] A. Khodaei, "Microgrid Optimal Scheduling With Multi-Period Islanding Constraints," *IEEE Trans. Power Syst.*, vol. 29, no. 3, pp. 1383–1392, May 2014.
[3] A. Khodaei and M. Shahidehpour, "Microgrid-Based Co-Optimization of Generation and Transmission Planning in Power Systems," *IEEE Trans. Power Syst.*, vol. 28, no. 2, pp. 1–9, 2012.
[4] S. Bahramirad, A. Khodaei, J. Svachula, and J. R. Aguero, "Building Resilient Integrated Grids: One neighborhood at a time.," *IEEE Electrif. Mag.*, vol. 3, no. 1, pp. 48–55, Mar. 2015.
[5] A. Khodaei, "Provisional Microgrids," *IEEE Trans. Smart Grid*, vol. 6, no. 3, pp. 1107–1115, 2015.
[6] A. Khodaei, "Resiliency-Oriented Microgrid Optimal Scheduling," *IEEE Trans. Smart Grid*, vol. 5, no. 4, pp. 1584–1591, Jul. 2014.
[7] A. Khodaei, S. Bahramirad, and M. Shahidehpour, "Microgrid Planning Under Uncertainty," *IEEE Trans. Power Syst.*, vol. 30, no. 5, pp. 2417–2425, 2015.
[8] A. Ipakchi and F. Albuyeh, "Grid of the future," *IEEE Power Energy Mag.*, vol. 7, no. 2, pp. 52–62, Mar. 2009.
[9] J. G. Kassakian, R. Schmalensee, G. Desgroseilliers, T. D. Heidel, K. Afridi, a M. Farid, J. M. Grochow, W. W. Hogan, H. D. Jacoby, J. L. Kirtley, and Others, "The future of the electric grid," Massachusetts Institute of Technology, Cambridge, MA, 2011.
[10] O. Zinaman, A. M. Miller, and D. Arent, *Power Systems of the Future*. NREL, 2015.
[11] J. Taft and A. Becker-Dippmann, "Grid Architecture," *PNNL*, 2015. [Online]. Available: http://energy.gov/sites/prod/files/2015/04/f22/QER Analysis - Grid Architecture_0.pdf.
[12] L. Kristov and P. De Martini, "21st Century Electric Distribution System Operations," *Caltech*. 2014.
[13] New York State Department of Public Service, "Developing the REV Market in New York: DPS Staff Straw Proposal on Track One Issues," 2014.
[14] P. De Martini and L. Kristov, "Distribution Systems in a High Distributed Energy Resources Future: Planning, Market Design, Operation and Oversight," no. 2. Lawrence Berkeley National Laboratory, 2015.
[15] E. Martinot, L. Kristov, and J. D. Erickson, "Distribution System Planning and Innovation for Distributed Energy Futures," *Curr. Sustain. Energy Reports*, vol. 2, no. 2, pp. 47–54, Apr. 2015.
[16] P. De Martini, "MORE THAN SMART: A Framework to Make the Distribution Grid More Open, Efficient and Resilient." Greentech Leadership Group, Aug-2014.
[17] F. Rahimi and S. Mokhtari, "From ISO to DSO: imagining new construct--an independent system operator for the distribution network," *Public Util. Fortn.*, vol. 152, no. 6, pp. 42–50, 2014.
[18] M. Keay, J. Rhys, and D. Robinson, "Distributed Generation and its Implications for the Utility Industry," in *Distributed Generation and its Implications for the Utility Industry*, Elsevier, 2014, pp. 165–187.
[19] Z. Wang, B. Chen, J. Wang, M. M. Begovic, and C. Chen, "Coordinated Energy Management of Networked Microgrids in Distribution Systems," *IEEE Trans. Smart Grid*, vol. 6, no. 1, pp. 45–53, Jan. 2015.
[20] J. Tong and J. Wellinghoff, "Rooftop Parity: Solar for Everyone, Including Utilities," *Public Util. Fortn.*, vol. 152, no. 7, pp. 18–23, 2014.
[21] "New York State Embarks On Bold New Vision," *Electr. J.*, vol. 27, no. 6, pp. 1–3, Jul. 2014.
[22] L. Kristov and D. Hou, "Distributed Generation and its Implications for the Utility Industry," in *Distributed Generation and its Implications for the Utility Industry*, Elsevier, 2014, pp. 359–377.
[23] S. Huang, Q. Wu, S. S. Oren, R. Li, and Z. Liu, "Distribution Locational Marginal Pricing Through Quadratic Programming for Congestion Management in Distribution Networks," *IEEE Trans. Power Syst.*, vol. 30, no. 4, pp. 2170–2178, Jul. 2015.
[24] R. Li, Q. Wu, and S. S. Oren, "Distribution Locational Marginal Pricing for Optimal Electric Vehicle Charging Management," *IEEE Trans. Power Syst.*, vol. 29, no. 1, pp. 203–211, Jan. 2014.
[25] S. Parhizi and A. Khodaei, "Market-based Microgrid Optimal Scheduling," in *IEEE SmartGridCommConf.*, 2015.
[26] S. Parhizi and A. Khodaei, "Investigating the Necessity of Distribution Markets in Accomodating High Penetration Microgrids," in *IEEE PES Transmission & Distribution Conference & Exposition*, 2016.
[27] S. Parhizi, A. Khodaei, and S. Bahramirad, "Distribution Market Clearing and Settelement," in *IEEE PES General Meeting*, 2016.
[28] W. H. Kersting, "Radial distribution test feeders," in *2001 IEEE Power Engineering Society Winter Meeting. Conference Proceedings (Cat. No.01CH37194)*, 2001, vol. 2, pp. 908–912.